\documentclass[sigconf]{acmart}
\usepackage{booktabs}  
\usepackage{siunitx}   
\usepackage{caption}   
\usepackage{multirow}     
\usepackage{subcaption} 
\usepackage{threeparttable}
\usepackage{url}
\usepackage{graphicx,amsmath}
\usepackage{tikz}
\usepackage{pgfplots}
\usepackage{pifont}
\usepackage[ruled,linesnumbered,noend]{algorithm2e}
\usepackage{algorithmic} 
\usepackage[normalem]{ulem} 
\AtBeginDocument{%
  }

\copyrightyear{2025}
\acmYear{2025}
\setcopyright{acmlicensed}\acmConference[CIKM '25]{Proceedings of the 34th
ACM International Conference on Information and Knowledge
Management}{November 10--14, 2025}{Seoul, Republic of Korea}
\acmBooktitle{Proceedings of the 34th ACM International Conference on
Information and Knowledge Management (CIKM '25), November 10--14, 2025,
Seoul, Republic of Korea}
\acmDOI{10.1145/3746252.3761266}
\acmISBN{979-8-4007-2040-6/2025/11}
\begin{document}

\title[ORCAS: Obfuscation-Resilient Binary Code Similarity Analysis using Dominance Enhanced Semantic Graph]%
{ORCAS: \texorpdfstring{\uline{O}}{O}bfuscation-\texorpdfstring{\uline{R}}{R}esilient Binary \texorpdfstring{\uline{C}}{C}ode Similarity \texorpdfstring{\uline{A}}{A}nalysis \\%
using Dominance Enhanced \texorpdfstring{\uline{S}}{S}emantic Graph}

\author{Yufeng Wang}
\email{wangyufeng2023@email.szu.edu.cn}
\orcid{0009-0000-2005-2172}
\affiliation{%
 \institution{College of Computer Science \\and Software Engineering, \\Shenzhen University}
 \city{Shenzhen}
 \state{Guangdong}
 \country{China}}

\author{Yuhong Feng}
\authornote{Corresponding author.}
\email{yuhongf@szu.edu.cn}
\orcid{0000-0002-7691-5587}
\affiliation{%
  \institution{College of Computer Science \\and Software Engineering, \\Shenzhen University}
  \city{Shenzhen}
  \state{Guangdong}
  \country{China}
  \postcode{518060}
}

\author{Yixuan Cao}
\email{caoyixuan2019@email.szu.edu.cn}
\orcid{0009-0006-6241-4251}
\affiliation{%
  \institution{College of Computer Science \\and Software Engineering, \\Shenzhen University}
  \city{Shenzhen}
  \state{Guangdong}
  \country{China}
}

\author{Haoran Li}
\email{lihaoran2018@email.szu.edu.cn}
\orcid{0009-0007-6789-5573}
\affiliation{%
  \institution{College of Computer Science \\and Software Engineering, \\Shenzhen University}
  \city{Shenzhen}
  \state{Guangdong}
  \country{China}
}

 \author{Haiyue Feng}
\email{fenghaiyue2018@email.szu.edu.cn}
\orcid{0009-0002-7239-3467}
\affiliation{%
 \institution{College of Computer Science \\and Software Engineering, \\Shenzhen University}
 \city{Shenzhen}
 \state{Guangdong}
 \country{China}}

\author{Yifeng Wang}
\email{wangyifeng2020@email.szu.edu.cn}
\orcid{0009-0006-4271-7378}
\affiliation{%
 \institution{College of Computer Science \\and Software Engineering, \\Shenzhen University}
 \city{Shenzhen}
 \state{Guangdong}
 \country{China}}

\renewcommand{\shortauthors}{Yufeng Wang et al.}

\begin{abstract}
Binary code similarity analysis (BCSA) serves as a foundational technique for binary analysis tasks such as vulnerability detection and malware identification. Existing graph based BCSA approaches capture more binary code semantics and demonstrate remarkable performance. However, when code obfuscation is applied, the unstable control flow structure degrades their performance. To address this issue, we develop ORCAS, an \underline{O}bfuscation-\underline{R}esilient B\underline{C}S\underline{A} model based on Dominance Enhanced \underline{S}emantic Graph (DESG). The DESG is an original binary code representation, capturing more binaries' implicit semantics without control flow structure, including inter-instruction relations (e.g., def-use), inter-basic block relations (i.e., dominance and post-dominance), and instruction-basic block relations. ORCAS takes binary functions from different obfuscation options, optimization levels, and instruction set architectures as input and scores their semantic similarity more robustly. Extensive experiments have been conducted on ORCAS against eight baseline approaches over the \textsc{BinKit} dataset. For example, ORCAS achieves an average 12.1\% PR-AUC improvement when using combined three obfuscation options compared to the state-of-the-art approaches.
In addition, an original obfuscated real-world vulnerability dataset has been constructed and released to facilitate a more comprehensive research on obfuscated binary code analysis. ORCAS outperforms the state-of-the-art approaches over this newly released real-world vulnerability dataset by up to a recall improvement of 43\%.

\end{abstract}




\maketitle

\section{Introduction}

{\em Code obfuscation} obscures the code's control flow and basic blocks while preserving the function semantics,
which has been widely applied to increase the difficulty of unauthorized reverse engineering \cite{gao2023obfuscation}, e.g., in 2018, roughly 50\% of popular apps on Google Play with over 10 million downloads were obfuscated to protect against plagiarism and repackaging \cite{wermke2018}. While malware authors also exploit code obfuscation techniques to conceal malicious contents and evade detection \cite{obfuscation2021,cuiying2024}. In 2024, approximately eight million new malware samples are identified, with over 300,000 being detected each day.\footnote{\url{https://portal.av-atlas.org/malware}}

Binary code similarity analysis (BCSA) scores the semantic similarity between two given binary code snippets, which is commonly used for malware identification~\cite{control2013,kim2019}, vulnerability detection~\cite{vulhawk2023,sense2023}, patch analysis~\cite{cross2016,spain2017,kargen2017}, and software plagiarism detection~\cite{semantics2017}, etc. Existing BCSA approaches perform well in BCSA tasks across different compilers, optimization levels, and instruction set architectures (ISAs). However, they face challenges when analyzing obfuscated binary code. To be specific, a binary function can be represented as a {\em control flow graph (CFG)}, where a node represents a basic block, and the directed edge between two nodes represents their control flow. 
There are three widely used code obfuscation techniques: {\em instruction substitution (SUB)}, {\em bogus control flow (BCF)}, and {\em control flow flattening (FLA)} \cite{obfuscator2015}. SUB obfuscation replaces simple instructions with semantically equivalent complex instructions, with control flow untouched. BCF obfuscation modifies the control flow by adding bogus basic blocks and edges. While FLA obfuscation complicates the control flow with a complex hierarchy, and then flattens the control flow into a linear form. Meanwhile, different techniques can be combined to obfuscate the code, e.g., for a given CFG, first, BCF is applied to double the number of nodes and edges, then FLA is applied to double them again \cite{asm2vec2019}. Then we can see that a CFG and its obfuscated CFG can have different nodes, different flow structures, and even arbitrarily added bogus nodes and edges, which present great challenges to existing BCSA approaches to detect their semantic similarity.

According to how the binary code is represented, existing BCSA approaches can be categorized into {\em instruction stream based} and {\em graph based}.
When instruction stream based approaches are used, 
binary code is first translated into assembly code or intermediate representation (IR) code, which are treated as natural language sentences. Then natural language processing (NLP) technologies are applied to capture the code's implicit semantics. NLP approaches have strong contextual understanding capabilities, however, they usually rely on pre-training tasks tailored to binary code, which require large-scale datasets and expensive computational resources \cite{jtrans2022,palmtree2021,binshot2022,order2020,vulhawk2023}. In addition, obfuscating simple instructions with complex ones increases contextual complexity and token length, which degrades the performance of NLP technologies.

\begin{figure*}[tb]
    \centering
    \subfloat[A CFG $G$\label{fig:motivation(a)}]{\includegraphics[page=1, height=0.18\textwidth]{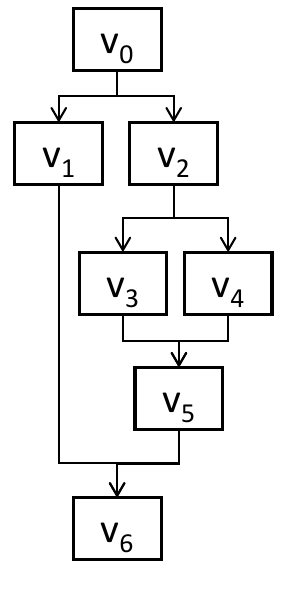}}
    \hfill
    \subfloat[$G$'s D-Tree\label{fig:motivation(b)}]{\includegraphics[page=2, height=0.18\textwidth]{figures/motivation_before_and_after_v4.pdf}}
        \hfill
    \subfloat[$G$'s PD-Tree\label{fig:motivation(c)}]{\includegraphics[page=3, height=0.1833\textwidth]{figures/motivation_before_and_after_v4.pdf}}
        \hfill
    \subfloat[$\Tilde{G}$: The Obfuscated $G$\label{fig:motivation(d)}]{\includegraphics[page=4, height=0.18\textwidth]{figures/motivation_before_and_after_v4.pdf}}
        \hfill
    \subfloat[$\Tilde{G}$'s D-Tree\label{fig:motivation(e)}]{\includegraphics[page=5, height=0.18\textwidth]{figures/motivation_before_and_after_v4.pdf}}
        \hfill
    \subfloat[$\Tilde{G}$'s PD-Tree\label{fig:motivation(f)}]{\includegraphics[page=6, height=0.18\textwidth]{figures/motivation_before_and_after_v4.pdf}}
    \caption{The D-tree and PD-tree of the CFG before and after the application of the BCF and FLA obfuscation, where gray nodes denote fake basic blocks generated by the obfuscation techniques. }
    \label{fig:motivation}
\end{figure*}

When graph based approaches are applied, a binary function is represented as a feature enhanced CFG or {\em semantics-oriented graph (SOG)}. Then graph neural network (GNN) technologies are used to learn the code's semantics. First, various features are incorporated into the CFG for capturing code's semantics: (1) Incorporating basic block features (e.g., number of instructions), which are obtained by manually crafting \cite{gemini2017,genius2016} or automatically generating using NLP technologies \cite{vulhawk2023,order2020,jia2024cross}; (2) Incorporating data flow graph (DFG) and call graph (CG) to provide richer semantic information between instructions and functions \cite{semantics2024,deepbindiff2020,guo2022}. Second, when a binary function is represented as a SOG \cite{sog2024}, nodes represent instruction opcodes and operands and edges reveal intra-instruction as well as inter-instruction relations, where control flow between basic blocks are represented using inter-instruction relations. Graph based approaches capture more binary code semantics and demonstrate remarkable BCSA performance. However, when code obfuscation technologies are applied, the unstable control flow structure degrades graph based BCSA performance \cite{asm2vec2019}, this problem remains unresolved.

This paper makes 4 main contributions for obtaining obfuscation-resilient BCSA:
\begin{itemize}
    \item Propose an original Dominance Enhanced Semantic Graph (DESG) embedding for representing a binary function, removing the control flow structure while revealing inter-basic block relations using dominance and post-dominance relations with a constructed dominator tree, and meanwhile revealing other graph semantics of existing BCSA approaches.  
    
    \item Develop ORCAS, a novel Obfuscation-Resilient BCSA model, capturing more binaries' implicit semantics without control flow structure, for more robust BCSA.
    
    \item Construct an original obfuscated real-world vulnerability dataset to evaluation of 1-day vulnerability detection on different obfuscation techniques.
    
    \item Conduct extensive experiments to evaluate the performance of ORCAS against baselines over the existing dataset and our constructed dataset, and experimental results demonstrate the effectiveness of ORCAS.

\end{itemize}


\section{Motivation}\label{sec:Background}
This section describes the motivation behind this paper. In general, BCSA can be conducted across different granularity, ranging from coarse to fine: program-level, function-level, slice-level, basic block-level, and instruction-level \cite{ASurveyofBinaryCodeSimilarity2021}. This paper focus on the function-level BCSA since functions are considered basic building blocks of program functionality, function-level BCSAs~\cite{jtrans2022,xu2023improving,vulhawk2023,sog2024,mvulpreter2022,asm2vec2019,Trex2023,crabs-former2024} enables many real-world applications such as patch analysis~\cite{PATCHDETECTOR2019, Robin2023}, and can be easily extended to detect similarities between whole programs.

Actually, a binary function can be represented as a CFG, denoted as $G=(\mathcal{V},\mathcal{E})$, where \(\mathcal{V} = \{v_0,v_1,\ldots,v_{n-1}\}\) is the set of nodes, \(v_0\) and \(v_{n-1}\) denote the unique entry and exit node respectively. $\mathcal{E} = \{(v_i, v_j) \mid v_i, v_j \in  \mathcal{V}\}$ is the set of directed edges between nodes. To be specific, a node represents a basic block, which is a sequence of consecutive instructions ending with a branch or return instruction.

To put the discussion into perspective, an example CFG $G$ is depicted in Fig.~\ref{fig:motivation(a)} to illustrate the problem. SUB obfuscates instructions, i.e., the nodes in a CFG, while FLA and BCF obfuscate nodes as well as edges in a CFG, which are widely adopted by various BCSA approaches \cite{sog2024,palmtree2021,guo2022,similarity2017,codeextract2024,vulhawk2023} for more notable performances. Then, BCF and FLA obfuscation have been applied to CFG $G$ and obtained an obfuscated CFG $\Tilde{G}$, which is depicted in Fig.~\ref{fig:motivation(d)}, where the dashed boxes display a typical application of an official Obfuscator-LLVM's BCF and FLA obfuscation techniques \cite{obfuscator2015}. We can see that node \(v_3\) and node \(v_4\) have an edge to node \(v_5\) in CFG $G$, while in the obfuscated CFG $\Tilde{G}$, neither node \(v_3\) nor node \(v_4\) has an edge to node \(v_5\). That is, the control flow structure of CFG $G$ is different from that of CFG $\Tilde{G}$.

From a CFG, a {\em dominator tree} ({\em D-Tree}) and a {\em post-dominator tree} ({\em PD-Tree}) can be constructed to capture the dominance and post-dominance relations respectively. Figs. \ref{fig:motivation(b)} and \ref{fig:motivation(c)} depicts the D-Tree and PD-Tree constructed upon CFG $G$. Generally, in a D-Tree,
node \(v_i\) dominates node \(v_j\) if and only if node \(v_i\) is included in every path from the entry node \(v_0\) to node \(v_j\), in this case node \(v_i\)
is the dominator of node \(v_j\).
Similarly, in a PD-Tree, 
node \(v_j\) post-dominates node \(v_i\) if and only if node \(v_j\) is included in every path from the node \(v_i\) to the exit node \(v_{n-1}\), in
this case node \(v_j\) is the post-dominator of node \(v_i\).
The dominance relations help identify the key nodes from the entry node to a specific node, while the post-dominance relations help identify the key nodes from a specific node to the exit node. 

\begin{table}[t]
\footnotesize
  \centering
  \caption{Average GED across different sizes of basic blocks.}
    \begin{tabular}{ccccc}
    \toprule
    \textbf{Basic Block Size Range} &
      \multicolumn{1}{l}{(0,50]} &
      \multicolumn{1}{l}{(50,100]} &
      \multicolumn{1}{l}{(100,150]} &
      \multicolumn{1}{l}{>150}
      \\
      \midrule
    \textbf{CFG}&
      207 &
      553 &
      916 &
      1084
      \\
    \textbf{D-Tree} &
      162 &
      431 &
      734 &
      828
      \\
      \bottomrule
    \end{tabular}%
  \label{tab:ged}%
\end{table}%

Figs. \ref{fig:motivation(e)} and \ref{fig:motivation(f)} depict $\Tilde{G}$'s D-Tree and PD-Tree respectively. We can see that there exists a dominance relation where node \(v_0\) dominates nodes \(v_6,v_2,v_1\), and node \(v_2\) dominates nodes \(v_3,v_4,v_5\) in the D-Tree of CFG $G$, and such dominance relations for node \(v_0\) and node \(v_2\) remain unchanged in the D-Tree of CFG $\Tilde{G}$ in Fig. \ref{fig:motivation(e)}. Similarly, there are post-dominance relations from node \(v_6\) to nodes \(v_1,v_5,v_0\), and from node \(v_5\) to nodes \(v_2,v_3,v_4\) in the PD-Tree of CFG $G$, and we can also have such post-dominance relations retain in the PD-Tree of CFG $\Tilde{G}$ in Fig. \ref{fig:motivation(f)}.

Based on the aforementioned discussion, we observe that dominance and post-dominance relations exhibit superior structural stability compared to control flow structure under obfuscation. To quantify this stability difference, we employ graph edit distance (GED), similar to \cite{ged2016}, which measures the minimum number of edit operations (i.e., node/edge additions/deletions or node relabelling) required to transform one graph into another. Specifically, we compute GED between original and obfuscated versions of both CFGs and D-Trees using approximately 1,800 function samples obfuscated with combined BCF and FLA obfuscation. Table \ref{tab:ged} presents a comparative analysis of structural stability between CFG and D-Tree under obfuscation across sizes of basic blocks. Notably, D-Trees exhibit a 23.6\% lower average GED for functions with basic block sizes exceeding 150. This demonstrates that D-Tree consistently exhibit superior structural stability compared to CFG under obfuscation.

\section{ORCAS}
\label{sec:ESG for BCSA}
\begin{figure*}[tb]
    \centering
    \includegraphics[width=\textwidth]{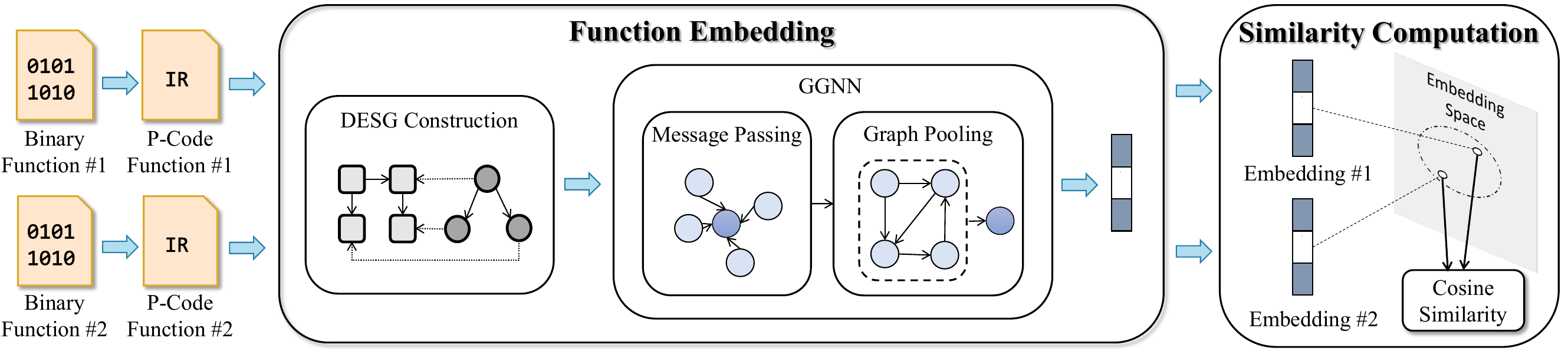}

    \caption{The framework of ORCAS. It first translates binary functions into  P-Code functions, generates function embeddings, and then computes their cosine similarity.}
    \label{fig:framework}

\end{figure*}

This section elaborates on how ORCAS is designed. 
Fig. \ref{fig:framework} depicts the ORCAS model, which takes two binary functions as input and decompiles them using Ghidra \cite{ghidra2020} into P-Code IR functions. In the function embedding phase, we construct the DESGs and employ a Gated Graph Neural Network (GGNN)~\cite{GGNN2015} to generate embeddings. 
In the similarity computation phase, we utilize cosine similarity to measure binary function pairs' similarity using their embeddings.

\subsection{Intermediate Representation}
\label{sec:Intermediate_Representation}

To mitigate the differences between various ISAs and simplify instruction types, IRs are widely utilized to represent binary functions in recent BCSA approaches \cite{cfg2vec2023,sog2024,vulhawk2023}. Since Ghidra outperforms other decompilers like IDA-Pro \cite{ida2011} and Angr \cite{angr2016} in \textit{correctly identified function starts} task \cite{disco2021}, we utilize Ghidra to decompile the binary code and lift it into P-Code, Ghidra’s IR.

According to reference manual\footnote{\url{https://spinsel.dev/assets/2020-06-17-ghidra-brainfuck-processor-1/ghidra_docs/language_spec/html/pcoderef.html}}, 
P-Code is a register transfer language designed for reverse engineering applications.
It translates processor instructions into a sequence of P-Code \textit{operations},
each P-Code operation takes one or more \textit{varnodes} as input and may produce a single output varnode. A varnode is a generalized abstraction of a register or memory location, represented by a formal triple: (\textit{address space type}, \textit{offset}, 
\textit{size}), where the offset denotes the displacement within a specific type of address space, and the size defines the allocated byte length. 

The address space for P-Code is a generalization of random access memory and is defined simply as an indexed sequence of bytes that can be read and written by the P-Code operations. There are five types of address spaces: \textit{ram}, \textit{register}, \textit{constant}, \textit{unique}, and \textit{stack}. Ram address space represents an abstraction of the physical memory accessed by the processor, while register address space refers to the processor's general-purpose registers. Unique address space holds intermediate value generated during processing, and stack address space denotes the memory space accessed through the stack pointer. Finally, constant address space is reserved for storing fixed values.

\begin{table}[tbp]
  \centering
  \caption{An example of varnode normalization strategies.}
  \footnotesize{
    \begin{tabular}{ccc}
    \toprule 
    \textbf{Address Space} & \textbf{Before Normalization} & \textbf{After Normalization} \\
    \midrule
    Ram & \texttt{(ram,0x8,8)} & \texttt{ram} or \texttt{abs} \\
    Register &\texttt{(register,0x0,4)} & \texttt{x86\_r\_0} or \texttt{ARM\_r\_0} \\
    Constant & \texttt{(const,0x0,4)} & 
    \texttt{c\_0} \\
    Unique &  \texttt{(unique,0x0,4)} & \texttt{unique} \\
    Stack & \texttt{(stack,0xf,8)} & \texttt{stack} \\
      \bottomrule 
    \end{tabular}%
    }
  \label{tab:varnodeNormal}%
\end{table}%

\textbf{Varnode Normalization.} Since each binary function contains too many different varnodes (e.g., with various offsets), it is difficult for the GNN's vocabulary to encompass all varnodes. To tackle this issue, we propose varnode normalization strategies, as outlined in Table \ref{tab:varnodeNormal}. We first disregard the size of all varnodes and then apply tailored strategies for different address spaces to reduce the vocabulary size: 
\begin{itemize}
    \item For ram address space, if the varnode represents a standard library function, we normalize it using the function name; 
    if not, we normalize it with the identifier \texttt{ram}. Notably, standard library function retain their original function name by default under obfuscation, as symbol table-based dynamic linking would fail with obfuscated name~\cite{picheta2020}.  
    \item For register address space, due to the varying number of registers across ISAs, we retain the offset while incorporating an architecture identifier (i.e., \texttt{ARM} or \texttt{x86}).
    \item For constant address space, we retain the offset and normalize the varnode with the identifier \texttt{c\_offset}.
    \item For unique and stack address spaces, we normalize them to the identifiers \texttt{unique} and \texttt{stack}, respectively.
\end{itemize}

\subsection{DESG Construction}
\label{sec:Enhanced_semantic_Graph}

To capture the comprehensive and stable semantics of the obfuscated binary function, we propose a graph based representation called the DESG. 
We first discuss the implicit semantics of the binary function, and then elaborate on how the DESG is constructed.

\subsubsection{Implicit Semantics of Binary Function} 
Different to natural language, binary code has well-defined implicit semantics, which can be categorized into \textit{inter-basic block relations}, \textit{inter-instruction relations}, and \textit{instruction-basic block relations}.

\textbf{Inter-basic block relations.} Control flow relations define the original relations between basic blocks in a CFG. However, bogus control flow generated by obfuscation techniques may introduce misleading information. As a result, we chose to remove them. Based on CFG, we introduce two new relations between basic blocks: dominance and post-dominance relations, derived from dominance analysis~\cite{lengauer1979fast,azen2003dominance}. These relations are crucial for understanding control flow structure and analyzing the execution paths of obfuscated binary function.

\textbf{Inter-instruction relations.} We further divide inter-instruction relations into two categories: \textit{data relation} and \textit{effect relation}.
In previous binary analysis works \cite{palmtree2021,similarity2017,sog2024,guo2022,codeextract2024,vulhawk2023}, the def-use relation between instructions is often used since it is a fundamental concept in program analysis. We adopt the def-use relation as the data relation, which reveals how instruction use values are defined by others. Inspired by \cite{click1995simple,sog2024}, effect relation imposes restrictions on the execution order between instructions, which is derived from def-use relations and it is essentially a special case of the def-use relation. Especially the effect relation focuses on the execution order between a \texttt{CALL} instruction and an instruction that may read from or write to memory.

\textbf{Instruction-basic block relations.} A CFG inherently implies that instructions are contained within specific basic blocks. To explicitly expose this to the GNN, we introduce a \textit{contain relation} between basic blocks and instructions, indicating that an instruction belongs to a particular basic block. This relation directly reduces the distance between instructions within the same basic block in the graph, facilitating information interaction within the GNN.

\begin{figure*}[ht]
    \centering
    \centering
    \includegraphics[width=\textwidth]{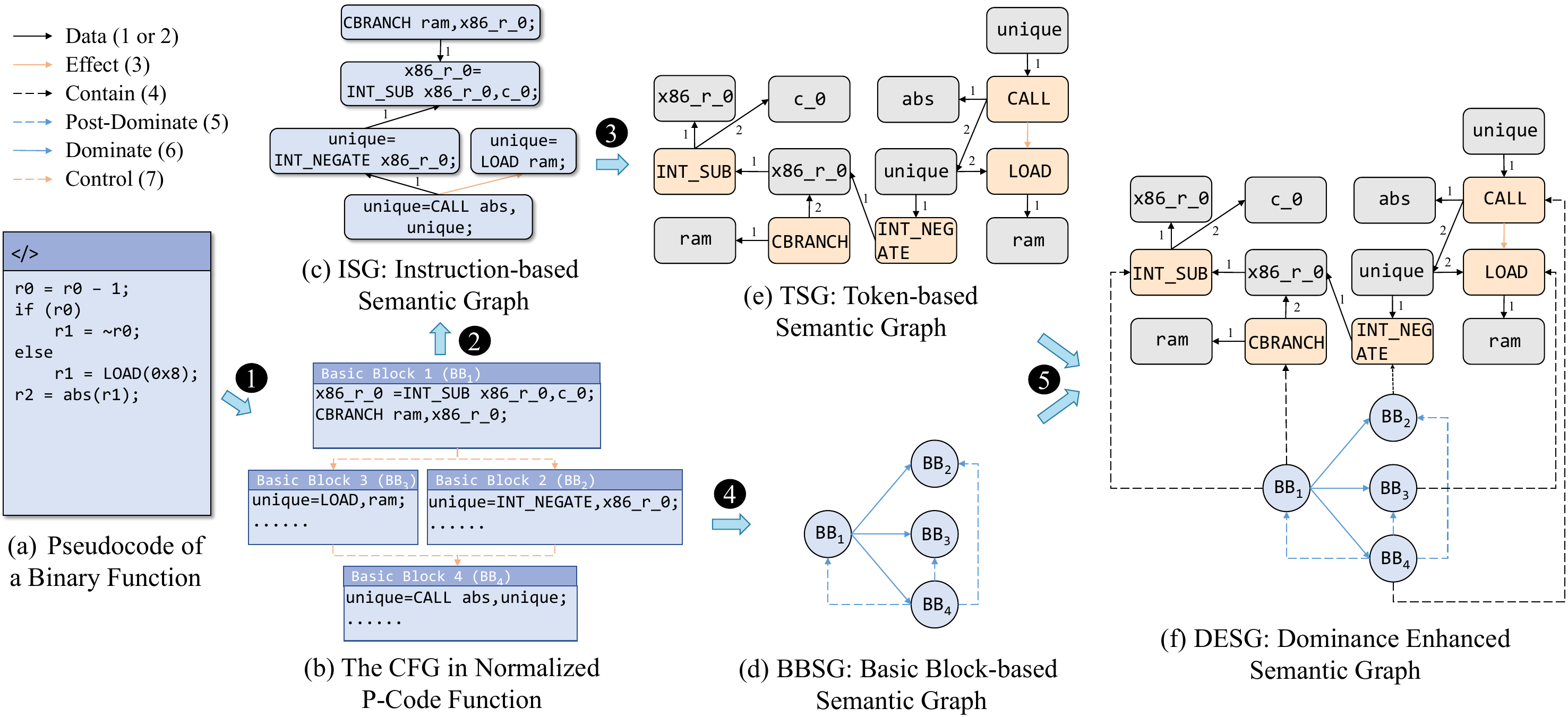}
    \caption{An example of constructing the DESG. For clarity, each basic block in (b) shows only the core logic P-Code operations. }
    \label{fig:graphConstruction}
\end{figure*}

\subsubsection{Steps to the DESG} Fig. \ref{fig:graphConstruction} presents the construction process of the DESG, which can intuitively be divided into five steps. The details are outlined in Algorithm \ref{alg:the Construction of ESG}. 

\noindent\textbf{Step \large\ding{182}. }We first decompile the binary function to obtain the CFG in the original P-Code, and then apply varnode normalization strategies to generate the CFG in the normalized P-Code function in Fig. \ref{fig:graphConstruction}b.

\noindent\textbf{Step \large\ding{183}. }We split the instruction sequence in each basic block into individual instructions, where a node represents an individual instruction, while an edge represents the relation between instructions (i.e., data relation or effect relation).
Thus, we obtain the instruction-based semantic graph (ISG).

\noindent\textbf{Step \large\ding{184}. }
Based on the ISG, we further reveal the structure within all instructions by splitting the opcode and operand(s) (i.e., varnodes) into tokens. As a result, the relations between instructions are refined into relations between tokens, with the effect relation specifically defined on the opcode tokens. Similar to \cite{sog2024}, the relation between opcode tokens and operand tokens can be considered a data relation, which we refine into two distinct relations: one for \textit{the first operand} and one for  \textit{the second operand} (if exists). 
Consequently, we obtain the token-based semantic graph (TSG).

\noindent\textbf{Step \large\ding{185}. }
We first create virtual basic block nodes for each basic block, and then introduce dominance and post-dominance relations between them by utilizing dominance analysis. These two relations not only cover the original control flow relations but also provide more stable semantic information than control flow relations.
For example, there is no directed edge between \texttt{BB}$_1$ and \texttt{BB}$_4$ in Fig. \ref{fig:graphConstruction}b, but in Fig. \ref{fig:graphConstruction}d, two directed 
edges are established between virtual basic block nodes \texttt{BB}$_1$ and  \texttt{BB}$_4$ due to their dominance and post-dominance relations. As a result, we construct the basic block-based semantic graph (BBSG).

\noindent\textbf{Step \large\ding{186}.} We introduce contain relation between the opcode token on the TSG and the virtual basic block node on the BBSG. A directed edge from a virtual basic block node to an opcode node if and only if the instruction of the opcode is contained within the corresponding basic block. Finally, we obtain the DESG.

In summary, the DESG includes three types of nodes and five types of edges. 
First, a node in a DESG can be one of the following types: (1) Virtual basic block node (e.g., \texttt{BB}$_1$), where we use the index of the basic block as the node attribute. (2) Opcode node (e.g., \texttt{INT\_SUB}), with its mnemonic serving as the node attribute. (3) Operand node (e.g., \texttt{x86\_r\_0}), where we use the token after varnode normalization strategies as the node attribute.
Second, five types of edges can be defined between nodes in a DESG: (1) Data relation, which reveals the node using values defined by another node. (2) Effect relation, which imposes the necessary execution order. (3) Contain relation, indicating that an opcode node is contained within a virtual basic block node. (4) Post-dominate relation, which reveals the post-dominance relations between virtual basic block nodes. (5) Dominate relation, which indicates the dominance relations between virtual basic block nodes. 

\begin{algorithm}[tbp]
\footnotesize
\caption{The construction algorithm of the DESG}
\label{alg:the Construction of ESG}
\KwIn{A binary function $f$.}
\KwOut{A DESG $\mathcal{G}_{\text{DESG}}$ for representing the binary function $f$.}
Let $\mathcal{G}_{\text{DESG}}$, $\mathcal{G}_{\text{ISG}}$,  and $\mathcal{G}_{\text{BBSG}}$ be empty graphs\;
Decompile $f$ into the CFG $\mathcal{G}_\text{CFG}$\;
Apply varnode normalization strategies on $\mathcal{G}_\text{CFG}$\;

\For{$block \in \mathcal{G}_\textnormal{CFG}$}{               
    \For{$inst \in \textnormal{GetInstruction}(block)$}{
        $\mathcal{G}_{\text{ISG}}$.\textnormal{addNode($inst$)}\;

        \For{$op \in \textnormal{the source operands of}\  
         inst$}{
           \If{$op$ \textnormal{is defined by instruction} $defInst$}{
                $\mathcal{G}_{\text{ISG}}$.\textnormal{addDataEdge($inst$, $defInst$)}\;
                    \If{$defInst$ \textnormal{accesses memory}}{
                        $\mathcal{G}_{\text{ISG}}$.\textnormal{addEffectEdge($inst$, $defInst$)}\;
                    }
           }
        }
    }
}
$\mathcal{G}_{\text{TSG}} \gets \mathcal{G}_{\text{ISG}}$\;
\For{$inst \in \mathcal{G}_{\textnormal{TSG}}$}{
    Split the opcode of $inst$ as a node $opc$\;
    Refine the effect edge of $inst$ into the effect edge of $opc$\;
    Split the target operand of $inst$ as a node $op$\;
    $\mathcal{G}_{\text{TSG}}$.\textnormal{addDataEdge($op$, $opc$)}\;
    \For{$op \in \textnormal{the source operands of}\  
    inst$}{
        Split $op$ as a node\;
        $\mathcal{G}_{\text{TSG}}$.\textnormal{addDataEdge($opc$, $op$)}\;
    }
}
$\mathcal{T}_\textnormal{D} \gets \textnormal{getDominatorTree}(\mathcal{G}_\text{CFG})$\;
$\mathcal{T}_\textnormal{PD} \gets \textnormal{getPostDominatorTree}(\mathcal{G}_\text{CFG})$\;
\For{$block \in \mathcal{G}_\textnormal{CFG}$}{
    Create a virtual basic block node $bb_i$ for $block$\;
    $\mathcal{G}_{\text{BBSG}}$.\textnormal{addNode($bb_i$)}\;

    $bb_j \gets \mathcal{T}_\textnormal{D}.\textnormal{getPredecessor}(bb_i)$\;
    $\mathcal{G}_{\text{BBSG}}$.\textnormal{addDomainceEdge($bb_j$, $bb_i$)}\;

    $bb_j \gets \mathcal{T}_\textnormal{PD}.\textnormal{getPredecessor}(bb_i)$\;
    
    $\mathcal{G}_{\text{BBSG}}$.\textnormal{addPostDominanceEdge($bb_j$, $bb_i$)}\;
}

Merge $\mathcal{G}_{\text{TSG}}$ and $\mathcal{G}_{\text{BBSG}}$ into $\mathcal{G}_{\text{DESG}}$ using contain edges\;
\end{algorithm}

\subsection{Model Training}
GGNN \cite{GGNN2015} is applied to generate embedding. We first present the GGNN's structure before discussing the model training. 
\label{sec:training}

\subsubsection{GGNN}
GNN is a neural network model for encoding nodes and edges, and learning their embeddings.
GNN consists of multiple \textit{message passing} layers and a \textit{graph pooling} layer.
GGNN is a variant of GNN, which extends GNN by incorporating a Gated Recurrent Unit (GRU) activation in the message passing phase to capture information from neighboring nodes. Following 
\cite{sog2024, sense2023, guo2022}, we adopt GGNN to learn the DESG's semantics.

\textbf{Message Passing.} A node updates its hidden state by aggregating the hidden states of its neighbors along with its state from the previous layer. Let $\mathcal{N}$ be the set of all nodes and $\mathbf{h}_u^l$ be the hidden state of node $u$ at the $l$-th layer of GGNN.
The aggregation of the neighbor's hidden states for node $u$ is formulated as Eqn.~\eqref{eq:hidden_state_update}.
\begin{equation}
\label{eq:hidden_state_update}
\footnotesize
\mathbf{m}_u^l = \sum_{v \in \text{Edge}(u)} f \left( \mathbf{h}_u^l,\mathbf{h}_v^l,\mathbf{e}_{v,u},\mathbf{e}_{u,v}\right)
\end{equation}
where $\text{Edge}(u)$ denotes the set of incoming and outgoing edges for node $u$, $\mathbf{e}_{u,v}$ represents the edge embedding of the direct edge $(u,v)$, and $f$ refers the application of two separate multilayer perceptrons used to aggregate incoming and outgoing edges, respectively. Then, in the $(l+1)$-th layer, the hidden state of node $u$ is formulated as Eqn.~\eqref{eq:h_cal}.
\begin{equation}
\label{eq:h_cal}
\footnotesize
\mathbf{h}_u^{l+1} = \text{GRU} \left( \mathbf{h}_u^{l},\mathbf{m}_u^l \right) 
\end{equation}
where GRU regulates information flow through gating mechanisms, retaining key features while discarding noise.

\textbf{Graph Pooling.}
After message passing layers, the final step is to pool all node embeddings into a graph embedding. Softmax \cite{softmax2020} is a widely used pooling method, which is formulated as Eqn.~\eqref{eq:softmax}.
\begin{equation}
\label{eq:softmax}
\footnotesize
\text{Softmax}(\mathcal{H} \mid \beta) = \sum_{\mathbf{h}_u^L \in \mathcal{H}} \frac{\exp \left( \beta  \odot \mathbf{h}_u^L \right)}{\sum_{\mathbf{h}_v^L \in \mathcal{H}} \exp \left( \beta \odot \mathbf{h}_v^L \right)} \odot \mathbf{h}_u^L
\end{equation}
where $\mathcal{H}$ refers to the set of all node hidden states and $\beta$ is a parameter known as the inverse temperature. Multi-head attention is a core technique in the transformer architecture \cite{vaswani2017attention}, allowing each head to learn different attention distributions. 
Similar to \cite{sog2024}, we integrate softmax pooling with multi-head attention, forming a stack of multiple softmax layers.
We first transform the hidden state $\mathbf{x}_u^k = \text{LayerNorm}\big(\text{ReLU}(\text{Linear}_k(\mathbf{h}_u^r))\big)$ using $k$ linear layers, followed by a Relu activation and layer normalization. Then, the calculation of every head is formulated as Eqn.~\eqref{eq:eg}.
\begin{equation}
\label{eq:eg}
\footnotesize
\mathbf{e}_g^k = \text{Softmax} \left( \left\{ \mathbf{x}_u^k \mid u \in \mathcal{N} \right\} \mid \beta^k \right) 
\end{equation}
where $k$ denotes the index of the head. Finally, all heads are concatenated and passed through a linear layer to produce the final graph embedding $\mathbf{e}_g$, formulated as $\mathbf{e}_g = \text{Linear}\big(\text{Concat}(\mathbf{e}_g^1,\mathbf{e}_g^2,\dots,\mathbf{e}_g^h)\big)$, where $h$ denotes the number of heads.

\subsubsection{Model Training} We employ distance weighted negative sampling \cite{wu2017sampling} with a margin-based pairwise loss \cite{GMN2019} to train the GGNN. This approach helps the model effectively distinguish between positive and negative samples, enhancing the capability of the embeddings. 
The loss function is formulated as Eqn.~\eqref{eq:loss}.
\begin{equation}
\label{eq:loss}
\footnotesize
\min_{\theta} \mathcal{L}(\theta) = \sum_{ \langle \mathbf{p},\mathbf{q} \rangle \in \mathcal{B}} \big( \text{positive}(\mathbf{p},\mathbf{q}) + \text{negative}(\mathbf{p}) \big)
\end{equation}
where $\mathcal{B}$ represents a training batch, consisting of a sequence of pairs, each in the form of 
$\langle \mathbf{p},\mathbf{q} \rangle$, with $\mathbf{p}$ and $\mathbf{q}$ denoting the graph embeddings originating from the same function. The positive computation is used in the loss function as $   \text{positive}(\mathbf{p},\mathbf{q}) = \max\big(1 - \text{cos}(\mathbf{p}, \mathbf{q}) - m, 0\big)$
and the negative sampling is applied as $\text{negative}(\mathbf{p}) = \max\big(\text{cos}(\mathbf{p}, \text{sampling}(\mathbf{p})) - m, 0 \big)$, 
where it samples a negative sample pair for $\mathbf{p}$ and calculates the cosine similarity, $m$ defines the margin that enforces a separation between positive and negative samples, ensuring positive pairs are closer and negative pairs are sufficiently separated.

\section{Experimental Setup}\label{sec:Experimental Setup}

\subsection{Dataset}
Our dataset is built on the \textsc{BinKit} dataset \cite{kim2022revisiting}, a large-scale benchmark for BCSA. It includes normal and obfuscation versions generated using the same optimization levels and ISAs. Normal versions include 3 ISAs (MIPS, ARM, and x86), 9 compiler versions (5 GCC, 4 Clang versions), and 5 different optimization levels (\texttt{O0} to \texttt{O3} and \texttt{Os}). Based on \textit{Obfuscator-LLVM (O-LLVM)} \cite{obfuscator2015}, obfuscation versions includes 4 obfuscation options: \textit{instruction substitution (SUB)}, \textit{bogus control flow (BCF)}, \textit{control flow flattening (FLA)}, and  a combination of all obfuscation options \textit{(SUB+BCF+FLA)}, referred to as \textit{ALL}.
To evaluate the effectiveness of ORCAS in analyzing the semantic similarity of binary functions across different obfuscation options, optimization levels, and ISAs, we integrate binaries from both normal and obfuscation versions. For our experiments, we select binaries from the x86-64 and ARM64 architectures, compiled with Clang for normal versions and O-LLVM (BCF, FLA, SUB, and ALL) for obfuscation versions, specifically with \texttt{O0} and \texttt{O3} optimization levels, covering 578 binaries.

\subsection{Metrics}
We choose several distinct metrics to comprehensively evaluate the performance of BCSA.

\subsubsection{One-to-One Similarity Matching}
Several works \cite{ASurveyofBinaryCodeSimilarity2021} utilize accuracy or recall to evaluate the performance of the model in similarity matching. However, these metrics at a fixed threshold may not fully reflect a model's ability to distinguish similar and dissimilar pairs.
Instead, we evaluate and compare the similarity matching performance of different approaches using the \textit{Precision-Recall (PR) curve}. 
Let $TP$, $FP$, and $FN$ represent the number of True Positive, False Positive, and False Negative binary function pairs inferred  by the model at a given threshold, respectively, we have $\text{Precision}={TP / (TP+FP)}$ and $\text{Recall}={TP / (TP+FN)}$.
To save space, we employ the \textit{the Area Under the PR Curve (PR-AUC)} as a comprehensive metric to summarize the overall performance, providing a more comprehensive comparison across different approaches.

\subsubsection{One-to-Many Similarity Searching}
Given a pool of original function 
\begin{math} 
    \{f_1,\dots, f_n\}
\end{math},
and a pool of target functions 
\begin{math} 
    \{g_1, \dots, g_n\}
\end{math}, where one of these target functions is 
\begin{math} 
    f_i^{gt}
\end{math}, corresponding to the group truth $f_i$.
The binary function similarity searching task aims to retrieve the top-k functions most similar to 
\begin{math} 
    f_i
\end{math} from target functions, ranked by their similarity score. The performance of the task is evaluated using \textit{Recall at top-k (Recall@k)} and \textit{the Mean Reciprocal Rank (MRR)}. Recall@\textit{k} measures the model's ability to include the group truth function within top-k results, which can be calculated by Eqn.~\eqref{eq:recall}.
\begin{equation}
\label{eq:recall}
\footnotesize
\text{Recall}@k={1\over n}\sum_{i=1}^n{\mathbb{I}\big({\text{Rank}(f_i^{gt})} \leq k\big)}
\end{equation}
where the indicator function $\mathbb{I}(x)$ maps boolean values to binary integers, with 0 for False and 1 for True.
MRR evaluates the model based on the rank of its first correct answer, as specified in Eqn.~\eqref{eq:mrr}.
\begin{equation}
\label{eq:mrr}
\footnotesize
    \text{MRR}={1\over n}\sum_{i=1}^{n}{1\over {\text{Rank}(f_i^{gt})}}
\end{equation}

\subsection{Baselines}
We compare ORCAS to the following eight baselines.

\textbf{SAFE \cite{SAFE2022}} employs a recurrent neural network
(RNN) augmented with attention mechanisms. It generates the
embeddings of binary functions from the assembly instructions.

\textbf{Gemini \cite{gemini2017}} takes the CFG as input, assigning manually crafted features to each basic block to construct an attributed CFG. It then uses GNN to calculate the similarity between binary functions by introducing a Siamese network.

\textbf{Asm2vec \cite{asm2vec2019}} uses random walks model the CFG as multiple sequences and applies the PV-DM model \cite{le2014distributed} to learn the embeddings of binary function.

\textbf{Graph Matching Networks (GMN) \cite{GMN2019}} proposes a variant of the GNN that incorporates a cross-graph attention-based matching mechanism. It takes pairs of CFG as input and computes a similarity score between them.

\textbf{Trex \cite{Trex2023}} first employs a hierarchical transformer-based model to capture the execution semantics of binary functions from micro traces. Then, it transfers the acquired semantic knowledge to match semantically similar binary functions. Our
implementation is directly based on the source code released by
the authors.\footnote{\url{https://github.com/CUMLSec/trex}} 

\textbf{jTrans \cite{jtrans2022}} proposes a jump-aware transformer-based model pre-trained on a large dataset, BinaryCorp, comprising approximately 21 million x86 binary functions. We fine-tune the pre-trained model released by the anthors on our dataset.\footnote{\url{https://github.com/vul337/jTrans}}

\textbf{CRABS-former \cite{crabs-former2024}} is a BCSA model built on top of jTrans. It introduces several assembly normalization strategies for x86-64 and ARM64 to improve the performance of BCSA. We implement CRABS-former and fine-tune it on our dataset. 

\textbf{HermesSim \cite{sog2024}} introduces novel SOG to represent binary function because it has well-defined semantic structures. Then, it utilizes GGNN to generate the embeddings of binary functions and calculate their cosine similarity. Our implementation is based on the source code released by the authors.\footnote{\url{https://github.com/NSSL-SJTU/HermesSim}}

Our implementation of SAFE, Gemini, Asm2vec, and GMN is based on the implementation provided by  \cite{marcelli2022machine}, with the default parameter settings.\footnote{\url{https://github.com/Cisco-Talos/binary_function_similarity}}

\section{Evaluation}\label{sec:Evaluation}
This section aims to address the following research questions (RQs).
\begin{itemize}
    \item \textbf{RQ1:} How performance is ORCAS on one-to-many similarity searching for binary functions compared with other baselines? (\S\ref{sec:RQ1:Searching against All Binary Functions})
    \item \textbf{RQ2:} How accurate is ORCAS on one-to-one similarity matching of obfuscated binary function compared with other baselines? (\S\ref{sec:Matching_Obfuscated_Function})
    \item \textbf{RQ3:} How performance is the impact of incorporating dominance analysis on the performance of ORCAS? (\S\ref{RQ3: Ablation Experiment})
    \item \textbf{RQ4:} How effective is ORCAS for real-world vulnerability detection compared to other baselines? (\S\ref{RQ4: Real-World Vulnerability Search})
\end{itemize}
We conduct all training and inference experiments on an x86 server running Ubuntu 20.04 LTS, equipped with an Intel Xeon 32-core 2.90GHz CPU, 256GB RAM,
and 2 NVIDIA GeForce RTX 3090 GPUs.

\subsection{RQ1: Searching against all Binary Functions}\label{sec:RQ1:Searching against All Binary Functions}

\label{sec:performance}
We evaluate the performance of ORCAS in searching similar binary functions across different obfuscation options, optimization levels, and ISAs for a given query function.
Specifically, we evaluate ORCAS through three BCSA searching subtasks: (1) cross-ISA (XA), (2) cross-optimization level (XO), and (3) cross-ISA, optimization level, and obfuscation option (XM). Notably, obfuscated binary functions are present in all three subtasks.
We conduct these evaluations on our dataset, setting the pool size for these subtasks to range from 10 to 2,000. 
Due to Trex's slow sampling rate, evaluation is limited to pool sizes below 1,000. Thus, we obtain the Recall@1 and MRR for different pool size settings, as shown in Tables \ref{tab:mrr_rec_xa}, \ref{tab:mrr_rec_xo}, and \ref{tab:mrr_rec_xm}.

In the XA searching subtask, ORCAS achieves 39.8\%/41.7\% higher Recall@1/MRR than CRABS-former when pool size is set to 2,000 via its 
architecture-agnostic IR design, while surpassing SOTA approach, HermesSim, by 9.6\%/7.4\%. Notably, Asm2vec's lack of cross-architecture support precluded evaluation in this subtask. 
In the XO searching subtask, ORCAS achieves 12.8\%/11.0\% improvement Recall@1/MRR, when pool size is set to 2,000, compared to HermesSim. 
In the XM searching subtask, the Recall@1/MRR of ORCAS surpass HermesSim by 9.6\%/8.4\% at a pool size of 2,000.

\begin{table}[bp]
\setlength{\tabcolsep}{1pt}
\footnotesize
    \centering
    \caption{Results of different BCSA approaches on XA subtask.}
    \begin{threeparttable}  
    \begin{tabular}
    {ccccccccc}
    \toprule
    \multicolumn{1}{c}{\multirow{2}{*}{Models}}  & \multicolumn{2}{c}{Pool size = 10} & \multicolumn{2}{c}{Pool size = 100} & \multicolumn{2}{c}{Pool size = 1,000} & \multicolumn{2}{c}{Pool size = 2,000} \\
    \cmidrule(r){2-3} \cmidrule(lr){4-5} \cmidrule(lr){6-7} \cmidrule(l){8-9}
    &{Recall@1} & {MRR} & {Recall@1} & {MRR} &{Recall@1} & {MRR} & {Recall@1} & {MRR} \\
    \midrule
    \multicolumn{1}{c}{SAFE}&0.506 & 0.685 & 0.133 & 0.269 & 0.023 & 0.069 & 0.011 & 0.042 \\
    \multicolumn{1}{c}{Gemini}&0.741 & 0.845 & 0.367 & 0.516 & 0.133 & 0.232 & 0.099 & 0.173 \\
    Asm2vec & - & - & - & - & - & - & - 
& - \\
    \multicolumn{1}{c}{GMN}&0.788 & 0.879 & 0.326 & 0.509 & 0.069 & 0.168 & 0.040 & 0.108 \\
    \multicolumn{1}{c}{Trex}&0.455 & 0.594 & 0.193 & 0.305 & - & - & - & - \\
    \multicolumn{1}{c}{jTrans}&0.467 & 0.668 & 0.089 & 0.225 & 0.014 & 0.049 & 0.008 & 0.030 \\
    \multicolumn{1}{c}{CRABS-former}&0.836 & 0.904 & 0.549 & 0.671 & 0.308 & 0.412 & 0.242 & 0.337 \\
    \multicolumn{1}{c}{HermesSim}&0.948 & 0.968 & 0.870 & 0.909 & 0.654 & 0.761 & 0.544 & 0.680 \\
    \midrule
    \multicolumn{1}{c}{ORCAS} & \textbf{0.961} & \textbf{0.974} & \textbf{0.913} & \textbf{0.937} & \textbf{0.744} & \textbf{0.825} & \textbf{0.640} & \textbf{0.754} \\
    \bottomrule
  \end{tabular}
         \end{threeparttable}
 \label{tab:mrr_rec_xa}
\end{table}

\begin{table}[tbp]
\footnotesize
\setlength{\tabcolsep}{1pt}
    \centering
    \caption{Results of different BCSA approaches on XO subtask.}
    \begin{threeparttable} 
    \begin{tabular}
    {ccccccccc}
    \toprule
    \multicolumn{1}{c}{\multirow{2}{*}{Models}}  & \multicolumn{2}{c}{Pool size = 10} & \multicolumn{2}{c}{Pool size = 100} & \multicolumn{2}{c}{Pool size = 1,000} & \multicolumn{2}{c}{Pool size = 2,000} \\
    \cmidrule(r){2-3} \cmidrule(lr){4-5} \cmidrule(lr){6-7} \cmidrule(l){8-9}
    &{Recall@1} & {MRR} & {Recall@1} & {MRR} &{Recall@1} & {MRR} & {Recall@1} & {MRR} \\
    \midrule
    \multicolumn{1}{c}{SAFE}&0.405 & 0.613 & 0.085 & 0.210 & 0.010 & 0.044 & 0.007 & 0.027 \\
    \multicolumn{1}{c}{Gemini}&0.527 & 0.681 & 0.144 & 0.274 & 0.033 & 0.080 & 0.015 & 0.048 \\
    \multicolumn{1}{c}{Asm2vec}&0.433 & 0.646 & 0.100 & 0.210 & 0.034 & 0.065 & 0.022 & 0.046 \\
    \multicolumn{1}{c}{GMN}&0.610 & 0.755 & 0.197 & 0.359 & 0.032 & 0.100 & 0.020 & 0.064 \\
    \multicolumn{1}{c}{Trex}&0.387 & 0.532 & 0.122 & 0.211 & - & - & - & - \\
    \multicolumn{1}{c}{jTrans}&0.401 & 0.604 & 0.111 & 0.224 & 0.019 & 0.060 & 0.031 & 0.060 \\
    \multicolumn{1}{c}{CRABS-former}&0.624 & 0.761 & 0.293 & 0.424 & 0.120 & 0.201 & 0.085 & 0.147 \\
    \multicolumn{1}{c}{HermesSim}&0.900 & 0.934 & 0.788 & 0.842 & 0.579 & 0.677 & 0.476 & 0.595 \\
    \midrule
    \multicolumn{1}{c}{ORCAS}&\textbf{0.937} & \textbf{0.957} & \textbf{0.860} & \textbf{0.896} & \textbf{0.690} & \textbf{0.769} & \textbf{0.604} & \textbf{0.705} \\
    \bottomrule
  \end{tabular}
    \end{threeparttable}
  \label{tab:mrr_rec_xo}
\end{table}

\begin{table}[tbp]
\setlength{\tabcolsep}{1pt}
\footnotesize
    \centering
    \caption{Results of different BCSA approaches on XM subtask.}
    \begin{threeparttable} 
    \begin{tabular}{
    ccccccccc
    }
    \toprule
    \multicolumn{1}{c}{\multirow{2}{*}{Models}}  & \multicolumn{2}{c}{Pool size = 10} & \multicolumn{2}{c}{Pool size = 100} & \multicolumn{2}{c}{Pool size = 1,000} & \multicolumn{2}{c}{Pool size = 2,000} \\
    \cmidrule(r){2-3} \cmidrule(lr){4-5} \cmidrule(lr){6-7} \cmidrule(l){8-9}
    &{Recall@1} & {MRR} & {Recall@1} & {MRR} &{Recall@1} & {MRR} & {Recall@1} & {MRR} \\
    \midrule
    \multicolumn{1}{c}{SAFE}&0.510 & 0.685 & 0.123 & 0.252 & 0.030 & 0.070 & 0.016 & 0.045 \\
    \multicolumn{1}{c}{Gemini}&0.473 & 0.646 & 0.157 & 0.275 & 0.058 & 0.103 & 0.046 & 0.078 \\
    Asm2vec & - & - & - & - & - & - & - & - \\
    \multicolumn{1}{c}{GMN}&0.726 & 0.832 & 0.319 & 0.481 & 0.055 & 0.150 & 0.032 & 0.097 \\
    \multicolumn{1}{c}{Trex}&0.302 & 0.460 & 0.154 & 0.240 & - & - & - & - \\
    \multicolumn{1}{c}{jTrans}&0.277 & 0.511 & 0.036 & 0.128 & 0.003 & 0.019 & 0.001 & 0.013 \\
    \multicolumn{1}{c}{CRABS-former}&0.509 & 0.682 & 0.160 & 0.292 & 0.036 & 0.089 & 0.027 & 0.069 \\
    \multicolumn{1}{c}{HermesSim}&0.908 & 0.939 & 0.816 & 0.859 & 0.642 & 0.723 & 0.557 & 0.660 \\
    \midrule
    \multicolumn{1}{c}{ORCAS}&\textbf{0.946} & \textbf{0.963} & \textbf{0.878} & \textbf{0.907} & \textbf{0.737} & \textbf{0.804} & \textbf{0.653} & \textbf{0.744} \\
    \bottomrule
  \end{tabular}
  \end{threeparttable}
  \label{tab:mrr_rec_xm}
\end{table}

\subsection{RQ2: Matching Obfuscated Binary Functions}
\label{sec:Matching_Obfuscated_Function}

\begin{table*}[tbp]
\setlength{\tabcolsep}{1.5pt}
\footnotesize
    \centering
    \caption{PR-AUC socres of matching against obfuscated binary functions on different obfuscation techniques.}
    \begin{tabular}{ccccccccccccccccccccc}
    \toprule
    \multirow{2}{*}{Models}  & \multicolumn{5}{c}{Bogus Control Flow (BCF)} & \multicolumn{5}{c}{Control Flow Flattening (FLA)} &
    \multicolumn{5}{c}{Instruction Substitution (SUB)} & \multicolumn{5}{c}{SUB+BCF+FLA (ALL)} \\
    \cmidrule(lr){2-6} \cmidrule(lr){7-11} \cmidrule(lr){12-16} \cmidrule(lr){17-21}
    & Readline & Sed & Sharutils & Tar & Avg. & Readline & Sed & Sharutils & Tar & Avg. & Readline & Sed & Sharutils & Tar & Avg. & Readline & Sed & Sharutils & Tar & Avg. \\
    \midrule
    SAFE      & 0.072 & 0.079 & 0.095 & 0.078 & 0.081 & 0.080 & 0.094 & 0.075 & 0.074 & 0.080 & 0.142 & 0.142 & 0.153 & 0.153 & 0.147 & 0.054 & 0.059 & 0.055 & 0.046 & 0.053 \\
    Gemini    & 0.062 & 0.089 & 0.083 & 0.063 & 0.074 & 0.026 & 0.026 & 0.025 & 0.022 & 0.024 &0.515 & 0.557 & 0.570 & 0.548 & 0.547 & 0.018 & 0.020 & 0.017 & 0.016 & 0.017 \\
    Asm2vec  & 0.013 & 0.021 & 0.028 & 0.027 & 0.022 & 0.011 & 0.012 & 0.013 & 0.010 & 0.011 &0.292 & 0.422 & 0.371 & 0.499 & 0.396 & 0.007 & 0.007 & 0.007 & 0.007 & 0.007 \\
    GMN      & 0.329 & 0.282 & 0.359 & 0.301 & 0.317 & 0.165 & 0.180 & 0.229 & 0.184 & 0.189 &0.513 & 0.488 & 0.604 & 0.529 & 0.533 & 0.164 & 0.137 & 0.184 & 0.092 & 0.144 \\
    jTrans   & 0.055 & 0.067 & 0.126 & 0.108 & 0.089 & 0.063 & 0.071 & 0.100 & 0.097 & 0.082 &0.373 & 0.459 & 0.524 & 0.570 & 0.481 & 0.015 & 0.022 & 0.020 & 0.027 & 0.021 \\
    CRABS-former & 0.040 & 0.070 & 0.057 & 0.088 & 0.063 & 0.015 & 0.016 & 0.012 & 0.028 & 0.017 &0.620 & 0.686 & 0.755 & 0.772 & 0.708 & 0.014 & 0.014 & 0.011 & 0.011 & 0.012  \\
    HermesSim & 0.805 & 0.892 & 0.905 & 0.913 & 0.878 & 0.827 & 0.914 & 0.910 & 0.917 & 0.892 &0.908 & 0.925 & 0.981 & 0.959 & 0.943 & 0.684 & 0.850 & 0.823 & 0.699 & 0.764 \\
    \midrule
    ORCAS    & \textbf{0.840} & \textbf{0.935} & \textbf{0.965} & \textbf{0.946} & \textbf{0.921} & \textbf{0.881} & \textbf{0.967} & \textbf{0.971} & \textbf{0.939} & \textbf{0.939} & \textbf{0.947} & \textbf{0.957} & \textbf{0.993} & \textbf{0.969} & \textbf{0.966} & \textbf{0.792} & \textbf{0.935} & \textbf{0.933} & \textbf{0.800} & \textbf{0.865} \\
    \bottomrule
    \end{tabular}
   \label{tab:auc_score}
\end{table*} 

Similar to \cite{asm2vec2019,gtrans2024,Trex2023,crabs-former2024}, we evaluate ORCAS's ability to match obfuscated binary functions with their normal version. First, we select representative projects (i.e., \textit{Readline}, \textit{Sed}, \textit{Sharutils}, and \textit{Tar}) from our dataset. For each project, we compile all functions without applying any obfuscation techniques. Next, we compile the project again using the selected obfuscation technique. Then, we can obtain both normal and obfuscated versions for each function. We link the normal and obfuscated functions to establish a one-to-one match using function symbols for evaluation. We consider function pairs compiled from the same function as positive samples, with others treated as negative samples, setting a 1:100 ratio of positive to negative samples. Finally, we use the PR-AUC score as the evaluation metric for each project. 

Table \ref{tab:auc_score} presents the PR-AUC scores of ORCAS and other baselines on different obfuscation techniques. Notably, ORCAS outperforms HermesSim with average PR-AUC improvements of 4.3\%, 4.7\%, 2.3\%, and 12.1\% on BCF, FLA, SUB, and ALL obfuscation respectively.

\begin{figure}[b]
    \centering
    \includegraphics[width=.48\textwidth]{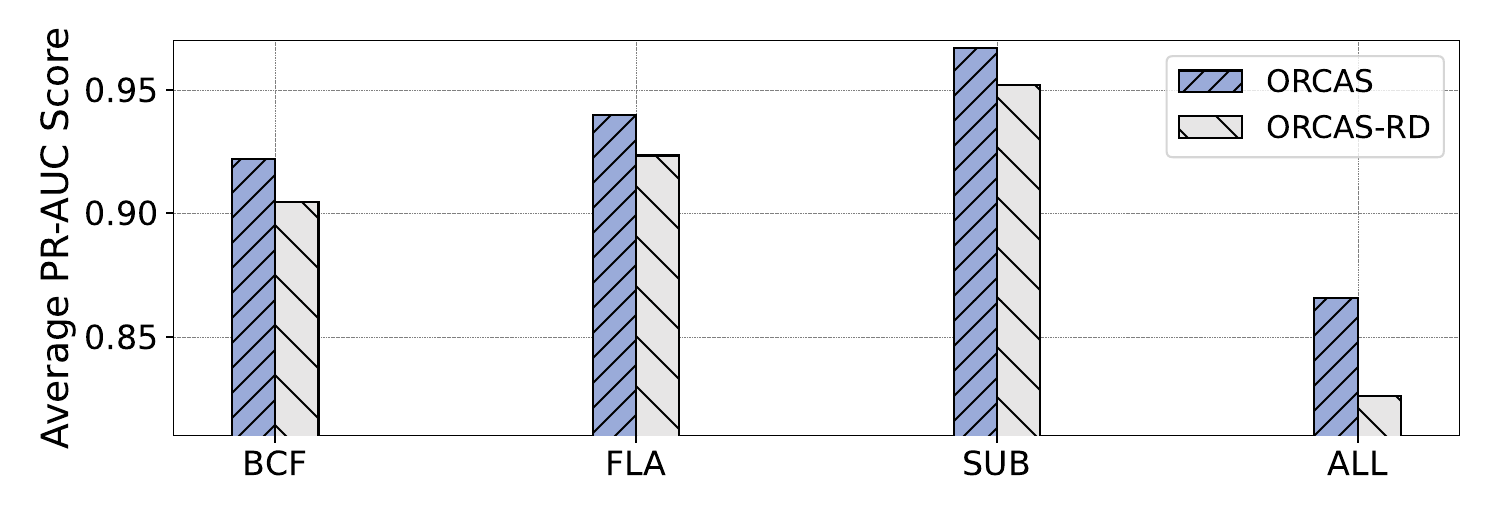}
    \caption{Average PR-AUC socres of ORCAS, {ORCAS-RD} on different obfuscation techniques. }
    \label{fig:ablation}
\end{figure}

\subsection{RQ3: Ablation Experiment}\label{RQ3: Ablation Experiment}
ORCAS employs a key technique by incorporating both dominance and post-dominance relations into the DESG. To demonstrate the effectiveness, we conduct ablation experiments using ORCAS-RD, a modified version that specifically removes the dominance and post-dominance relations from the DESG while retaining all other components. The experimental setup follows the same configuration as used in RQ2. 

Fig.~\ref{fig:ablation} presents a comparison of PR-AUC scores between ORCAS and ORCAS-RD. The results reveal a performance degradation on different obfuscation techniques.  Specifically, ORCAS-RD achieves average PR-AUC score degradations of 1.7\%, 1.7\%, 1.6\%, and 4.0\% on BCF, FLA, SUB and ALL obfuscation, respectively, compared to ORCAS.
The most significant performance gap occurs on ALL obfuscation, demonstrating that dominance analysis provides the greatest benefits when dealing with multiple combined obfuscation techniques.
Therefore, these results demonstrate that ORCAS's key technique effectively improves the accuracy and robustness of similarity detection between obfuscated binary functions.

\subsection{RQ4: Real-World Vulnerability Search}\label{RQ4: Real-World Vulnerability Search}
Previous works \cite{vulhawk2023,jtrans2022,cebin2024} utilize 1-day vulnerability detection to evaluate the effectiveness of BCSA approaches in real-world tasks. Similar to \cite{jtrans2022}, we perform our evaluation on eight Common Vulnerabilities and Exposures (CVEs) from a public dataset presented in \cite{wang2019detecting}.
We produce 10 variants for each function by O-LLVM \cite{obfuscator2015} with 2 ISAs (x86-64 and ARM64) and 5 obfuscation options (disable obfuscation, enable SUB, BCF, FLA, and ALL obfuscation options). We enable \texttt{-O3} optimization option to make them closer to the release version as suggested in \cite{kim2022revisiting}, and apply \texttt{-fno-inline-functions} option to facilitate locating a certain vulnerable function. Thus, we obtain a challenging obfuscated real-world vulnerability dataset, which has been released to facilitate future research on obfuscated binary code analysis.\footnote{\url{https://github.com/Cao-Wuhui/ORCAS}}

We provide details for evaluating the effectiveness of vulnerability detection. Similar to \cite{jtrans2022,cebin2024}, for each CVE, we first extract the functions in the 10 project variants to build a function pool $P$ and mark the 10 vulnerable functions in $P$. Then, we search in $P$ using a variant of the vulnerable function and it returns the top 10 similar binary functions. The number of function pools for each CVE varies from 18,719 to 57,673. Finally, assuming $m$ of the returned binary functions are marked, we calculate the recall rate by $m/10$. 

\begin{figure}[tb]
    \centering
    \includegraphics[width=.48\textwidth]{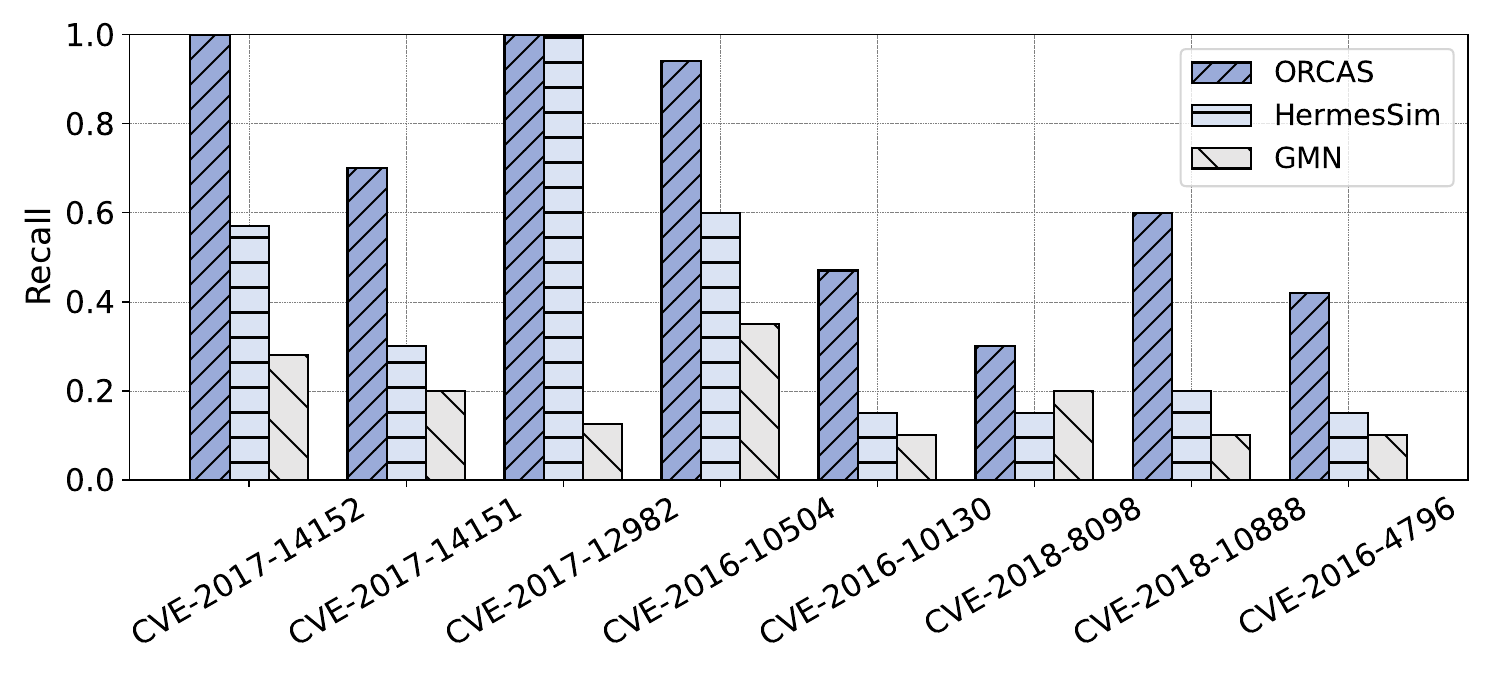}
    \caption{Recall rate of real-world vulnerability search. }
    \label{fig:real_world}
\end{figure}

In our experiment,  we select the vulnerable functions compiled with ALL obfuscation option as our queries, then use the average recall rate as the evaluation metric. Fig. \ref{fig:real_world} illustrates the results of the recall rate for each CVE. We compare ORCAS to two outstanding baselines: GMN and HermesSim. 
ORCAS outperforms both baseline models, e.g., for CVE-2017-14152, from the \texttt{openjpeg} project, which includes 18,719 functions, ORCAS achieved a 100\% recall rate, meaning it can recall all variants of the vulnerable functions, improving by 43\% over HermesSim and by 72\% over GMN. 
This demonstrates that ORCAS can be effectively applied to real-world vulnerability search.

\section{Related Work}\label{sec:related works}

This section provides a brief overview of recent BCSA approaches.

\subsection{Binary Code Representation in BCSA} Various representations are employed to capture the semantics of binary code, including trees, instruction streams, and graphs.

\textbf{Trees.} TEDEM~\cite{pewny2014leveraging} transforms basic block instructions into equations through symbolic simplification to generate expression trees. Recent approach by Asteria~\cite{asteria2021,asteria2024} finds that abstract syntax tree (AST) exhibit greater stability than CFG cross-ISA, employing Tree-LSTM networks to encode AST of a binary function into embeddings. RCFG2Vec~\cite{rcfg2vec2024} converts each assembly instruction into a three-level syntax tree, where the root represents the instruction itself and child nodes denote its opcode or operands, subsequently generating three types of embeddings.

\textbf{Instruction streams.}
Both assembly code and IR code are widely adopted for binary representation  \cite{SAFE2022,palmtree2021,jtrans2022,binshot2022,vulhawk2023}. Early approach SAFE~\cite{SAFE2022} employed skip-gram models, which predict surrounding instructions from a target instruction, to generate instruction embeddings. With advances in NLP technologies, pre-trained models have demonstrated superior performance. Researchers~\cite{palmtree2021,jtrans2022,binshot2022,vulhawk2023} develop specialized pre-training tasks to enhance semantic capture, with PalmTree~\cite{palmtree2021} implementing both context window prediction and def-use prediction tasks.
Moreover, Vulhark \cite{vulhawk2023} further enhancing IR through microcode-based optimizations including root-operand analysis and instruction simplification. However, substituting complex instructions for simpler ones significantly degrades representation effectiveness under obfuscation.

\textbf{Graphs.} 
CFG serve as a prevalent representation owing to their cross-ISA ability \cite{sog2024}, often extended through integration with other features. Early approaches~\cite{genius2016,gemini2017} construct a attributed CFG using manually designed basic block features, while recent approaches integrate CFG with CG and DFG~\cite{kim2022,deepbindiff2020,guo2022}. DeepBinDiff \cite{deepbindiff2020} merges CG and CFG to construct a inter-procedural CFG. 
XBA \cite{kim2022} propose a binary disassembly graph that leverages basic blocks and external functions. 
Additionally, HermesSim~\cite{sog2024} proposes a SOG that decomposes instructions into operand and opcode nodes to reveal their relations. 
These approaches achieve promising performance but become unstable under obfuscation due to their reliance on original control flow.

\subsection{Resilience to Obfuscation in BCSA}
Recent approaches have been addressing the challenges posed by obfuscation. Asm2vec \cite{asm2vec2019} applies random walk and edge coverage on the CFG to sample instruction sequences, then trains PV-DM model to generate function embeddings. CoP \cite{semantics2017} integrates program semantics with fuzzy matching of the longest common subsequence to capture basic block semantics. BinFinder \cite{binfinder2023} first identifies a set of engineered interpretable features, then trains a neural network model using a siamese network architecture. GTrans \cite{gtrans2024} leverages the D-Tree structure and employs three encoding strategies to train a graph attention network. HermesSim \cite{sog2024} introduces a SOG to represent binary functions and generates function embeddings using GGNN. Our experimental results shows its resilience against obfuscation. Trex \cite{Trex2023} learns program execution semantics from a mixture of regular and forced-execution traces. Although these approaches are resilient to obfuscation and can detect similarities between obfuscated binary, they still fall short of capturing the complete semantics of obfuscated binary, limiting their effectiveness when multiple obfuscation techniques are applied simultaneously.

\section{Conclusion}\label{sec:conclusion}
This paper proposes ORCAS, a novel
obfuscation-resilient BCSA model based on Dominance Enhanced Semantic Graph (DESG).
The DESG captures binaries' implicit semantics, including inter-basic block, inter-instructions and instruction-basic block relations, then gated graph neural networks are utilized to generate the embedding. 
We contribute the obfuscated real-world vulnerability dataset for evaluating 1-day vulnerability detection on different obfuscation techniques.
Extensive comparative experiments and ablation experiment demonstrate ORCAS's performance, showing that ORCAS outperforms baselines on both the \textsc{BinKit} dataset and our original obfuscated real-world vulnerability dataset.

\section*{GenAI Disclosure Statement}
No generative artificial intelligence (GenAI) tools are used in  any stage of the research.

\begin{acks}
\begin{sloppypar}
The authors would like to appreciate the anonymous reviewers for their feedback.
This project is supported by the Shenzhen Science and Technology Foundation (JCYJ20210324093212034), Guangdong Province Undergraduate University Quality Engineering Project (Shenzhen University Academic Affairs [2022] No. 7), National Natural Science Foundation of China (62272315), Science and Technology R\&D Program of Shenzhen (20220810135520002), Guangdong Basic and Applied Basic Research Foundation (2023A1515011296), the Stable Support Project of Shenzhen  (20231120145719001), Guangdong Province Key Laboratory of Popular High Performance Computers 2017B030314073, and Guangdong Province Engineering Center of China-made High Performance Data Computing System.
\end{sloppypar}
\end{acks}

\bibliographystyle{ACM-Reference-Format}
\bibliography{samples/ORCAS}










\end{document}